\documentclass[pop,aip,preprint]{revtex4}

\usepackage{siunitx}
\usepackage{setspace}

\usepackage[]{caption}
\usepackage[latin9]{inputenc}
\usepackage{xcolor}
\usepackage{verbatim}
\usepackage{graphicx}
\graphicspath{{./Figures/}} % Specifies the directory where pictures are stored
\PassOptionsToPackage{normalem}{ulem}
\usepackage{ulem}
\usepackage[section]{placeins}        %Keep Figure in Section
\usepackage{float}% If comment this, figure moves to Page 2
\usepackage{multirow}
\usepackage{epsfig}
\usepackage{pdfpages}
\usepackage{ragged2e}
\usepackage{amsmath, amsthm, amssymb, latexsym, nicefrac, xfrac, cancel}
\usepackage[colorlinks]{hyperref}
\usepackage[colorinlistoftodos]{todonotes}
\usepackage{tikz}

\makeatletter

\@ifundefined{textcolor}{}
{%
	\definecolor{BLACK}{gray}{0}
	\definecolor{WHITE}{gray}{1}
	\definecolor{RED}{rgb}{1,0,0}
	\definecolor{GREEN}{rgb}{0,1,0}
	\definecolor{BLUE}{rgb}{0,0,1}
	\definecolor{CYAN}{cmyk}{1,0,0,0}
	\definecolor{MAGENTA}{cmyk}{0,1,0,0}
	\definecolor{YELLOW}{cmyk}{0,0,1,0}
}

\newcommand{\oset}[3][0ex]{%
	\mathrel{\mathop{#3}\limits^{
			\vbox to#1{\kern-2\ex@
				\hbox{$\scriptstyle#2$}\vss}}}}
\makeatother

\makeatletter
\newcommand{\hvec}{\mathpalette{\overarrowsmall@\rightharpoonfill@}}
\def\rightharpoonfill@{\arrowfill@\relbar\relbar\rightharpoonup}
\newcommand{\overarrowsmall@}[3]{%
	\vbox{%
		\ialign{%
			##\crcr
			#1{\smaller@style{#2}}\crcr
			\noalign{\nointerlineskip}%
			$\m@th\hfil#2#3\hfil$\crcr
		}%
	}%
}
\def\smaller@style#1{%
	\ifx#1\displaystyle\scriptstyle\else
	\ifx#1\textstyle\scriptstyle\else
	\scriptscriptstyle
	\fi
	\fi
}
\makeatother

\newcommand{\be}{\begin{equation}}
\newcommand{\ee}{\end{equation}}

\newcommand{\bea}{\begin{eqnarray}}
\newcommand{\eea}{\end{eqnarray}}

\newcommand{\cmmnt}[1]{}

\makeatother

\begin{document}
	
\begin{comment}
Write your comments here!! This is how we comment in LaTex.
\end{comment}	

\title{A 3D Magnetohydrodynamic  simulation for the propagation of plasma plume transverse to applied magnetic field}

\author{Bhavesh G. Patel}
 \email{bhavesh@ipr.res.in}
\affiliation{ Institute for Plasma Research,  Bhat, Gandhinagar 382 428, India}

\author{ Narayan Behera, R. K. Singh,  Ajai Kumar}
\affiliation{ Institute for Plasma Research,  Bhat, Gandhinagar 382 428, India}
  
\author{Amita Das }
%\email{amita@iitd.ac.in}
\affiliation{Physics Department, Indian Institute of Technology, Delhi, Hauz Khas New Delhi -110016, India}
\date{\today}
 
\begin{abstract}
We have carried out a 3D ideal-MHD (Magnetohydrodynamic) simulation to study the evolution of laser generated plasma plume in a moderate external magnetic field (0.13 T) oriented perpendicular to the flow direction of the plasma plume. The simulation  shows that the plasma plume pushes the external magnetic field lines outward in the direction of the expansion. This leads to compression and bending of the magnetic field lines.The force resulting from the change in shape and the density of magnetic field lines opposes the expansion of the plume. An elliptic layer of shocked plasma is formed at the plasma/external field interface leaving a cavity in the plume core due to the outward expansion and the inertia of the plume. As the plasma pressure drops due to expansion, the imbalance between the magnetic energy and the internal energy results in the collapse of the cavity. These observations have striking similarities with the observations of the experiments [Phys. Plasmas 24, 033511 (2017)] performed recently to study the plasma plume expansion in the presence of an external transverse magnetic field. This similarity indicates that the physical mechanisms dominantly governing the plasma plume expansion in the moderate magnetic field are aptly described in the ideal MHD regime. The studies thus show that the laser generated plasma plume can be utilized to carry out interesting experiments on MHD phenomena in a simple laboratory set up.
\end{abstract}


\cmmnt{3D ideal MHD simulation, Laser-produced plasma, Topology of Magnetic field lines, Diamagnetic cavity}

\maketitle

\section{Introduction}
One often encounters the interaction of flowing plasma with an external magnetic in various astrophysical and laboratory scenarios. Supernova explosions in the interstellar medium, impingement of coronal mass in the planetary magnetosphere, artificial plasma release in magnetosphere and propagation of laser-generated plasma plume in an external magnetic field are just a few examples to cite \cite{ArtificialPlasmaReleaseMagnetosphere,EjectionSolarCoronalMass,SupernovaExplosions}. Besides the fundamental aspect, the fact that a multitude of applications like exploration of the astrophysical plasmas \cite{PlasmaApplicationAstroPhys0}, propulsion of space vehicles \cite{PlasmaApplicationPropulsionSpaceVehicles}, manipulation of plasma plume in the Pulsed Laser Deposition(PLD) \cite{PlasmaApplicationPLD}, increase in the sensitivity of Laser Induced Breakdown Spectroscopy (LIBS) \cite{PlasmaApplicationLIBS}, thermonuclear devices  \cite{PlasmaApplicationThermonuclearDevices}  etc.relies on the interaction of plasma with the magnetic field makes this an important topic of research.  The physical understanding of the processes underlying the expansion of plasma plume in the presence of  external magnetic field is, therefore, important. 

Several researchers have carried out experimental and theoretical investigations over the past many years in the field of laser-produced plasma plume expansion in the presence of external magnetic field \cite{BEHERA20152215,PLAinMagneticField2}. They have highlighted several important aspects like formation of diamagnetic cavity \cite{PlasmBfieldInteracDiamagneticCavity,DiamagneticCavity1}, structure formation\cite{PlasmBfieldInteracStructureFormation, BfieldTopology2b}, plasma plume oscillation\cite{ResistivePlasmoidPenetratingBfield}, edge instability, and studied the dependence of these phenomena on plasma parameter and magnetic field.While many simulation studies exists for plasma plume expansion in external magnetic field parallel to the direction of propagation of plasma plume, \cite{Ciardi2013,ParallelBfield2,ParallelBfield3,ParallelBfield4,ParallelBfield5} there are only a few 3D simulation studies pertaining to plasma plume propagation in transverse magnetic field \cite{DiamagneticCavityInstabilityMHD1}. 

To best of our knowledge, no 3D simulation based on MHD has been carried out in moderate magnetic field range ($< \text{1 T}$) at which most of the laboratory experiments are usually performed. In this manuscript we present results of 3-D simulation based on an ideal Magnetohydrodynamics model for the evolution of a laser produced plasma plume in moderate transverse external field. We provide a qualitative comparison with experimental observations and discuss the dynamics of plasma plume propagation in the transverse magnetic field. The paper has been organized as follows. We provide a brief description of the simulation set up and geometry along with the boundary conditions that have been employed in section II.  In section III the simulation results are presented along with the comparison with a recent experimental observations. Section IV contains the summary and conclusion.

\section{Simulation setup and boundary conditions}
We aim to model the evolution of the laser produced plasma plume against an external background magnetic field which is aligned perpendicular to plasma plume flow direction. In order to relate the simulation closely to the experimental scenario, we perform the simulation in two stages. The first stage comprises production of plasma by laser-target interaction and the second stage comprises evolution of plasma in the external magnetic field. During the production phase the laser energy heats the target, causing the melting and the vaporization and possible ionization of its material. The initial plasma formation is modeled using 1D lagrangian VLL (Virtual laser laboratory) code \cite{VLL_Code1}. VLL code, models interaction of laser pulse with the target material and account for laser light absorption, plume expansion, electron thermal conductivity, two-temperature effects, phase transitions, material spallation. The Lagrangian plasma profiles of  density, pressure and velocity are then used to construct inflow profiles for the MHD simulation.This inflow profiles are used as an input to the modular  Godunov-type  parallel  3D ideal MHD code PLUTO \cite{PLUTOcode}. In PLUTO code the integration of ideal MHD equations is accomplished using finite volume formalism which comprises three main steps:an interpolation routine followed by the solution of Riemann problems at zone edges and a final evolution stage. It numerically solves the following set of MHD equations:

\begin{equation}
\begin{array}{l}
\frac{\partial}{\partial t}\left(\begin{array}{c}
\rho\\
\rho\hvec{v}\\
E_t\\
\hvec{B}
\end{array}\right)
+
\hvec{ \nabla}\cdot\left(\begin{array}{c}
\rho\hvec{v}\\
\rho\hvec{v}\hvec{v}-\hvec{B}\hvec{B}+Ip_{t}\\
\left(E_t+Ip_{t}\right)\hvec{v}-\hvec{B}\left(\hvec{v}\cdot\hvec{B}\right)\\
\hvec{v}\hvec{B}-\hvec{B}\hvec{v}
\end{array}\right)^{T}
=
\left(\begin{array}{c}
0\\
0\\
0\\
0
\end{array}\right)

\end{array}
\end{equation}

Here $\rho$ is the mass density, $\rho\hvec{v}$ is the momentum density, $\hvec{v}$ is the fluid velocity, $p$ is the thermal pressure, $\hvec{B}$ is magnetic field and $E_t$ is the total energy density.

\begin{equation}
E_t=\rho \epsilon + \frac{1}{2}\rho v^2 + \frac{B^2}{2}
\end{equation}

. In addition, $\rho \epsilon $ is the internal energy density, and we use the ideal gas equation of state for closure$\;\left(\rho \epsilon = \frac{P}{\gamma -1}\right)$.The simulation is performed in Cartesian coordinate system. In this work we have used linear interpolation for reconstruction of primitive variable. An approximate HLL Riemann solver was used to compute the inter-cell flux and Vanleer limiter was used as a flux limiter. The solution to the MHD equation must satisfy the solenoidal constraint ($\nabla \cdot \overset{\rightharpoonup}{B} =0$) at all time steps. We have used Constrained Transport (CT) approach to ensure the solenoidal condition. In this method, the magnetic field is defined at face centers, the electric field used to update the magnetic field is defined at cell edges, and the other fluid variables are defined at cell centers. The Stokes theorem is then used directly to integrate the induction equation. This ensures the solenoidal character of the magnetic field down to machine precision. A second-order Runge-Kutta scheme was then used to advance the MHD equations in time. Further,  PLUTO uses non-dimensional units. In PLUTO one has to specify the units of density($\rho_0 \hspace{2mm} gr/cm^3$), length($L_0 \hspace{2mm} cm$) and velocity($v_0 \hspace{2mm}cm/s$).  All the other units are then derived from these fundamental units. Time, pressure and magnetic field are measured in units of $t_0(=L_0/v_0)$, $p_0(=\rho_0 v_0^2)$, and $B_0(=\sqrt{4\pi \rho_0 v_0^2})$ respectively. In other words the normalized  density, velocity, pressure and magnetic field are given by $\rho_{sim}=\frac{\rho_{cgs}}{\rho_0}$, $v_{sim}=\frac{v_{cgs}}{v_0}$, $p_{sim}=\frac{p_{cgs}}{p_0}$ and  $B_{sim}=\frac{B_{cgs}}{\rho_0}$ respectively.

\begin{figure}[htbp]
	\centerline{\includegraphics[width=11cm, height=8cm]{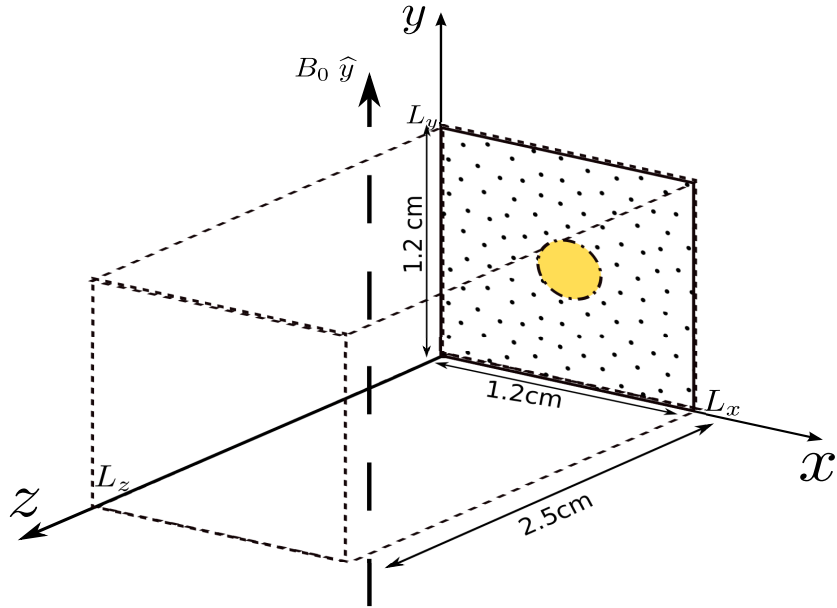}} %\centerline{\includegraphics[scale=0.5]{Figures/ComputationalDomain3D.eps}}
	\caption{Geometry for the computational domain.Outflow boundary condition is imposed on all planes confining the computational domain except the dotted plane with yellow circle. Inflow boundary condition is implemented in the yellow circle for about $8ns$ and then it is replaced by the reflective boundary condition. Reflective boundary condition is imposed on the dotted area.}
	\label{Fig0_ComputationalDomain3D}
\end{figure}

 One of the objectives of this simulation was to understand the findings of the experiment conducted at the Diagnostic division of Institute for Plasma research,India (IPR). In this experiment, Nd:YAG laser beam with pulse energy 150 mJ and pulse width 8 ns (FWHM)  was used to ablated 1 mm thick Aluminium plate. The focused laser beam diameter on the aluminium plate was $\sim 1mm$ which corresponds the laser fluence of $19.1 J/cm^2$. The target place was placed in between the Helmholtz coil which facilitate the plume expansion in uniform magnetic field. External magnetic field oriented in direction perpendicular to direction of propagation of the plasma plume was produced by Helmholtz coil along with capacitor bank-based pulse power system. Two mutually synchronised ICCD cameras mounted on orthogonal to the direction of plasma expansion has been used to record the time resolve images of the luminous plasma plume in the spectral range of $350-750 nm$. The magnetic field and exposure time of ICCD was set as $0.13 T$ and $5 ns$ respectively for the presented images in this report \cite{DiamagneticCavity3}.  The schematic in  Fig. \ref{Fig0_ComputationalDomain3D} shows the computational domain.In order to compare the experimental observations with the simulation results, we have chosen the extent of the computational domain in both $x \text{ and }y$ direction to be \SI{1.2}{cm} . The length of the computational domain in z directions is taken as \SI{2.5}{cm}. The grid size of \SI{0.0025}{cm}   was taken in all the directions.  As shown in the figure, the inflow takes place at the centre of x-y plane and the plasma propagates in z direction.  An external magnetic field of  \SI{ 0.13}{T} has been applied along the $\hat{y}$ direction. Except for the shaded region, reflective boundary condition is imposed on the xy plane at $z=0$. Inflow boundary condition  is imposed in the shaded region till the inflow persists (here it is for about \SI{8}{ns}). Further, for the VLL simulation,  we chose the target material to be Aluminum  and a Gaussian profile for the incident laser pulse. The intensity, wavelength and the pulse width of the laser pulse were  {\SI{2.24e9}{W/cm^2}}, {\SI{1.064}{\mu}m} and \SI{8}{ns} respectively.  This corresponded to a fluence of \SI{19.1}{J/cm^2}. The maximum value of the particle density, mass density, pressure and velocity of the  inflowing plasma plume as obtained from the VLL simulation are about \SI{3.3e19}{particles/cm^3}, \SI{1.5 e-3}{g/cm^{3}}, \SI{1.8e9}{Ba}, \text{ and } \SI{2e6}{cm/s}  respectively. These parameters are similar to the experimental parameters considered in \cite{DiamagneticCavity3}. Once the inflow stops, reflective boundary condition is imposed on the shaded/inflow region.

\pagebreak
\section{Simulation results}
Many researchers have experimentally investigated the dynamical and geometrical features of the laser-generated plasma evolving in the external transverse magnetic field \cite{DiamagneticCavity1,DiamagneticCavity2}. These experiments have demonstrated the formation and collapse of the diamagnetic cavity on the plane perpendicular to the applied magnetic field. In present work we have performed ideal MHD simulations to understand the role of magnetic field in the evolution of the plasma plume. We have chosen the simulation parameters in a way that the experimental findings of the Diagnostic division at IPR \cite{DiamagneticCavity3} could be compared with the simulation results.  We use characteristic velocity  $v=2e6 cm/s$, length scale $L=0.1 cm$ (which is equal to diameter of the spot from which the inflow occurs), pressure $P=1.8e9 Ba$ and plasma density $1.5e-3 g/cm^3$  to calculate the dimensionless Reynolds, magnetic Reynolds and Peclet numbers for the present simulation \cite{ParallelBfield4}. The values of the dimensionless Reynolds, magnetic Reynolds and Peclet numbers are  $2.6e3, 2.6, 10.26$ respectively. Each of these numbers exceed unity shows that the advective transport dominates over the diffusive transport. Further, the ion collisionless skin depth is about $0.02cm$ which is less than the length scale of the system($L=0.1 cm$). Hence,  We have ignored the effect of viscosity, thermal conductivity, and Hall term and employed pure  ideal MHD equations to simulate plasma plume evolution \cite{IdealMHDJustification0,IdealMHDJustification1,HallMHD1} and would show later that even with the absence of these effects, the experimental observations  of cavity formation and its collapse is well reproduced. 

\begin{figure}[htbp]
	\centerline{\includegraphics[width=11cm, height=8cm]{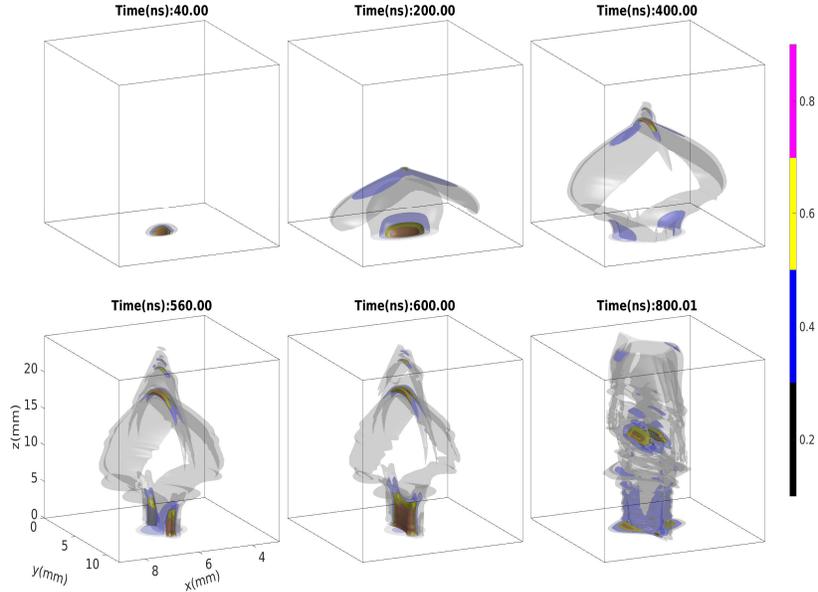}}
	\caption{Isosurface plots of the plasma plume density at different times. The figure displays four different normalized densities depicted in the color bar. Note different scale shown for x and y axes. }
	\label{Fig1_PlumeEvolution}
\end{figure}

 In  Fig.\ref{Fig1_PlumeEvolution} we have shown  a series 3D surface plots  of plasma density of the  plume at various times over a course of its  evolution. It can be observed that the evolution  follows two distinct phases. The plasma bulges out in the transverse direction in the first phase forming a magnetic cavity and in the second phase  the cavity  collapses.

\begin{figure}[htbp]
	\centerline{\includegraphics[width=11cm, height=8cm]{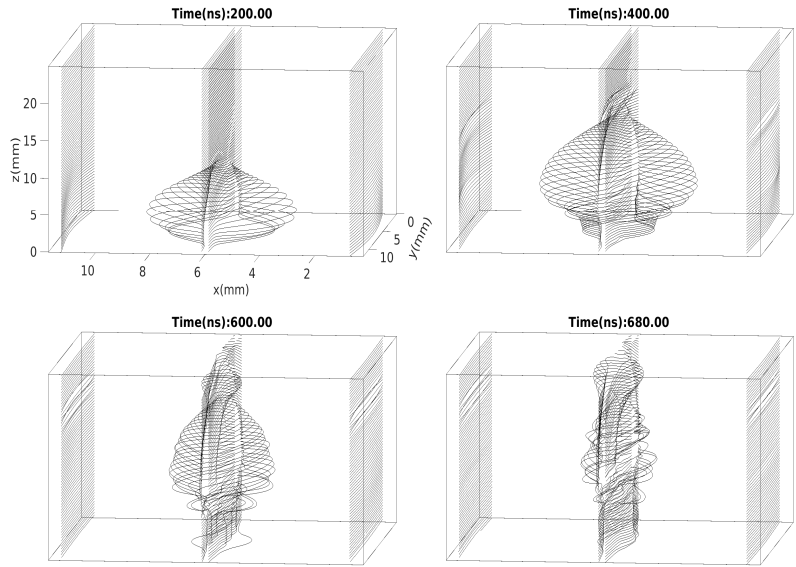}}
	\caption{The magnetic field lines at four different planes at different times.}
	\label{Fig2_BfieldTopology}
\end{figure}

This is also corroborated by the plot of magnetic field lines shown in Fig.\ref{Fig2_BfieldTopology} which shows the initial bulging of the magnetic field lines in the transverse direction. However, at a later stage the lines appear to collapse back. This can be understood by realizing that  initially $(0-80 ns)$ the plasma expands unabated under the effect of large thermal pressure which pushes the magnetic field lines from the high pressure region. The flow velocity of the plasma plume front is of the order of  \SI{1e6}{cm/s}. As time passes the plasma plume expands in all directions and a cavity starts appearing gradually in the plume. The cavity is observed to attain the maximum size at about \SI{300}{ns} and after this it begins to collapse. The change in the shape of the plasma plume observed in the Fig.\ref{Fig1_PlumeEvolution} is related to the change in the shape of the applied magnetic field and the thermal pressure of the plasma. Fig.\ref{Fig2_BfieldTopology} shows the shape of the external magnetic field lines at various times
We observe that as the plasma plume expands across the background magnetic field, it sweeps up the magnetic layers in accordance to the frozen-in condition in the ideal MHD framework. It is worth noting here that the plasma plume propagates at the velocity exceeding the Alfven velocity. If we use characteristic scale  $a=0.1cm$, magnetic field $B_0=1300 G$ and density $\rho_0=1.5e-3g/cm^3$, the  Alfven velocity turns out to be $V_A \sim  $\SI{9.5e3}{cm/s} and the Alfven time turns out to be $\tau_A \sim  $\SI{1e-5}{s}. For the present case the  Alfven Mach number($=V_{plumefront}/V_A$) is about $100$ . The high value of Alfven Mach number implies that the plasma expansion  governs the shape of the magnetic field for the initial time. The magnetic field lines advect with the plasma flow which results in bending and accumulation of field lines at the expanding plume front. The electromagnetic force density, $J \times B$, acting on the plasma plume due the change in the shape of the magnetic field could be understood in terms of lateral magnetic pressure $\nabla_{\perp}\left(\frac{B^2}{8\pi} \right)$ and the magnetic tension $\frac{B^2}{4\pi} $ acting along the lines of magnetic field. 

\cmmnt{
In order to understand the effect of the change in the topology of the magnetic field on the evolution of the plasma plume, consider the expression of the electromagnetic force density given below.
\begin{equation}
J \times B = -\nabla_{\perp}\left(\frac{B^2}{8\pi} \right) + \frac{B^2}{4\pi} \hat{\kappa}
\end{equation}
Here, $\hat{\kappa} $ is the magnetic field line curvature. The force $\frac{B^2}{4\pi} \hat{\kappa}$ can be attributed to the tension  $\frac{B^2}{4\pi} $ acting along the lines of force, while the force $\nabla_{\perp}\left(\frac{B^2}{8\pi} \right)$ is due to a lateral magnetic pressure. Thus, the $\vec{J}\times \vec{B} $ force acting on the plasma plume due to the curved  magnetic field lines produces  magnetic pressure and  tension  acting against the plasma pressure.

}

The in-homogeneity in the magnetic field leads to the force density pushing the field lines from the regions of high density to regions of low density in the direction perpendicular to the field. The bending and stretching of the magnetic field line leads to the magnetic tension. The magnetic field lines have a tendency to shorten and hence this force due to the magnetic tension points in a direction opposite to the direction of the convex curvature of the bent field lines. Once the plasma pressure drops as the result of expansion the magnetic field lines snap back.  The interplay between the magnetic field gradient, magnetic tension and the plasma pressure gradient govern the evolution of the plasma plume. The role of the magnetic tension and the magnetic pressure in governing the plasma plume structure is clearly evident from the Fig.\ref{Fig2_BfieldTopology} and Fig.\ref{Fig1_PlumeEvolution}.  The segregation and stretching of the magnetic field lines gives rise to  $\vec{J} \times \vec{B}$ force which opposes the expansion of plasma and there is increase in the magnetic field energy. 
  
  \begin{figure}[htbp]
  	\centerline{\includegraphics[width=11cm, height=8cm]{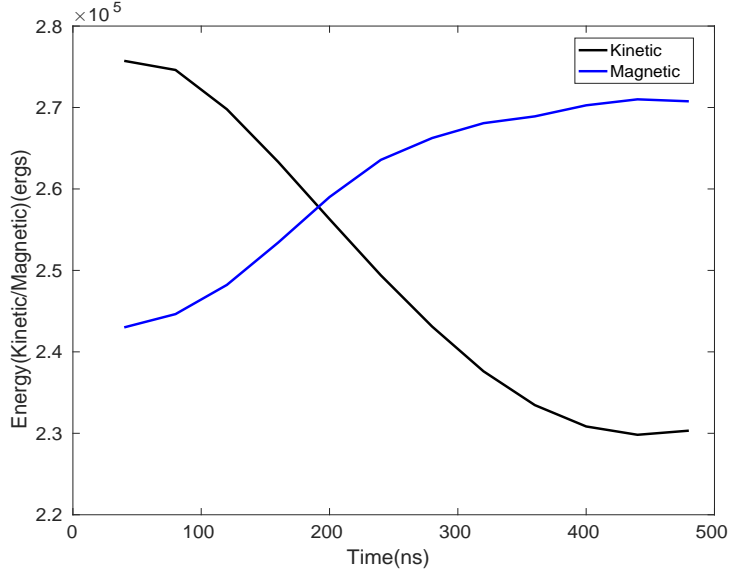}}
  	\caption{Kinetic and magnetic field energy as function of time.}
  	\label{Fig3_KEBE}
  \end{figure}

\begin{figure}[htbp]
	\centerline{\includegraphics[width=11cm, height=8cm]{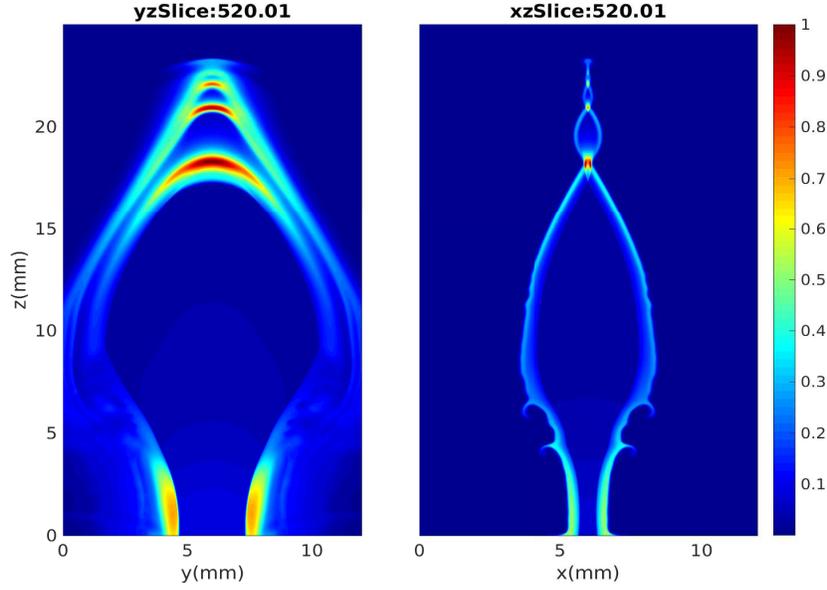}}
	\caption{The subplot on the left and right shows the cross-sectional view of the plasma plume density in the planes  parallel and perpendicular to the magnetic field respectively. The planes pass through the center of the plume.}
	\label{Fig4_2DParPerpCutPlane}
\end{figure}

In Fig.\ref{Fig3_KEBE} we show  the plot of plasma kinetic energy and magnetic field energy as function of time. The total energy which is the sum of these energies has also been plotted. Initially, both kinetic and magnetic energy (and thereby  total energy)  increases up to  \SI{8}{ns} as the plume enters the simulation box.Thereafter, till \SI{500}{ns} the total energy remains roughly constant. The small losses in total energy over the time  window shown in Fig. 4 could be due to slight numerical dissipation in the code. At \SI{500}{ns} the plasma plume hits the other boundary and starts moving out of the simulation box, causing a drop in total energy (not plotted in Fig. 4). In the intervening period of \SI{80 }{ns} to  \SI{500}{ns} for which the total energy is a constant, there is an exchange of energy between the kinetic and magnetic parts.  The kinetic energy associated with plasma decreases and the magnetic energy in the box increases. This happens as the plasma pushes the magnetic field in the transverse direction, thereby stretching the magnetic field lines as can be seen from Fig.\ref{Fig2_BfieldTopology}.  Thus, as plasma expands, its thermal pressure continuously decreases and the effect of $\vec{J} \times \vec{B}$ force dominates. As seen in the Fig.\ref{Fig1_PlumeEvolution}, the plasma from the core of the plume accumulates at the plume boundary thereby forming a cavity surrounded by a shell of shocked plasma. A distinct cavity begins to appear at around $120 ns$ and it reaches its maximum size at around $300ns$.  Further, it is observed  that as the plasma plume translates across the magnetic field, it expands without much deceleration in the direction of the applied magnetic field while the expansion of plasma experiences notable deceleration in the direction perpendicular to both magnetic field and the flow direction . This observation is well corroborated by the Fig.\ref{Fig4_2DParPerpCutPlane} which shows the pseudo-color plots of density in the cut-plane parallel and perpendicular to the direction of the applied magnetic field.  We also observe the fine structure  beyond \SI{15}{mm} that  resembles  an underexpanded supersonic jet. As the high pressure plasma plum inflows into the low pressure background, it is inhibited by the ambient gas and the external magnetic field which pushes the plume back and a converging conical shock is formed. The incident shock does not converge to a point on the axis and a Mach disk is formed in the front of the moving cocoon shaped cavity.

\begin{figure}[htbp]
	\centerline{\includegraphics[width=11cm, height=8cm]{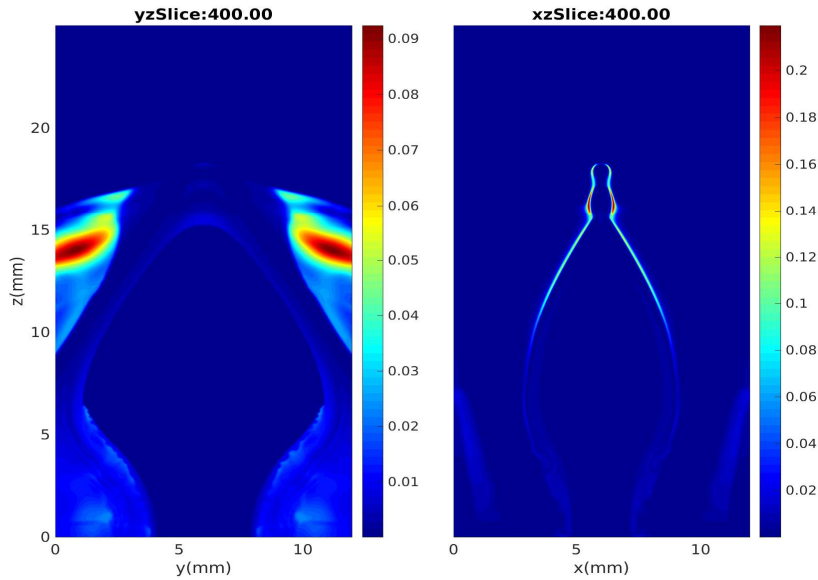}}
	\caption{Cross-sectional view of the plasma plume temperature (in kev)in the planes passing through center of plume and parallel and perpendicular to the externally applied magnetic field respectively.}
	\label{Fig5_CutPlanePlasmaTemperature}
\end{figure}

Further, the plasma plume is cooled initially due to free expansion and as the cavity begins to form the plasma shell is heated by the adiabatic compression at the field-plasma interface. This is depicted in Fig.\ref{Fig5_CutPlanePlasmaTemperature} where the cut-plane plot of the plasma temperature shows a bubble structure. 

The second phase of the plasma plume evolution commences once the cavity attains its maximum size. In this phase the plasma plume gradually begins to collapse under the action of $J \times B$ force. Since the plasma expands freely across the plane containing the magnetic field, the plasma plume assumes the shape of a flat pancake as depicted by the last subplot in the  Fig.\ref{Fig1_PlumeEvolution}. 
We now compare the results of the MHD simulation with the experimental observations made by the Diagnostic division at IPR \cite{DiamagneticCavity3}. In order to compare the diamagnetic cavity formed in the experiment and the simulation, we plot the images of the plasma plume obtained from the experiment and plume density obtained simulation together. It is important to note that the intensity of the ICCD images of plasma plume is commensurate with the density of emitting ionic and neutral species of the plasma plume.In fact, the intensity recorded by ICCD is sum total of the intensities coming from the planes constituting the plasma plume. 

\begin{figure}[htbp]
	\centerline{\includegraphics[width=11cm, height=8cm]{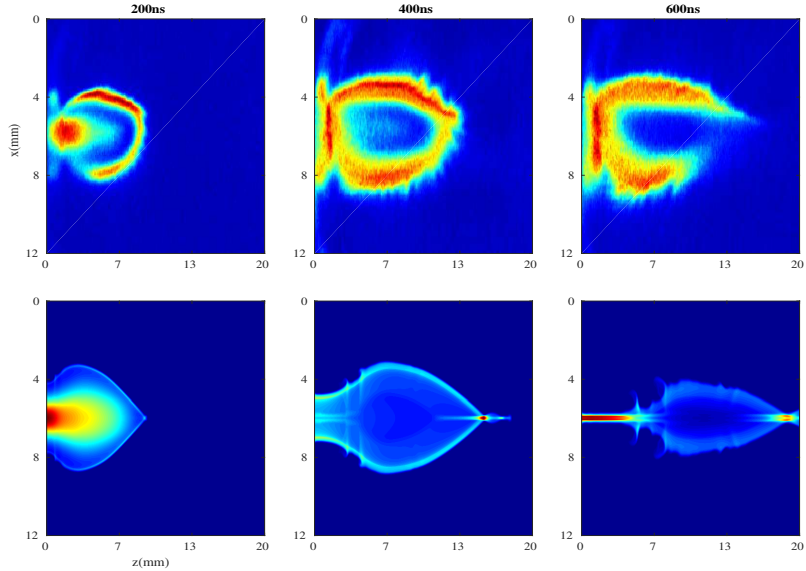}}
	\caption{ 
		Images of the plasma plume obtained from experiment and simulation at various times in the plane (x-z ) perpendicular to the direction of magnetic field $ (\hat{y})$.  The top row shows the experimental images while the bottom row shows simulation images obtained after adding all the layers of the  plume density perpendicular to the magnetic field.
	}
	\label{Fig6_SimVsExpt}
\end{figure}

The subplots of the top row of Fig.\ref{Fig6_SimVsExpt} shows the images of the expanding plasma plume obtained with an ICCD camera oriented in direction parallel to the applied magnetic field. The presented ICCD images are the spectrally integrated images in the range of $\sim 350-750 nm$.  The emissions from the plume species consists of the line emission from the plume species as Bremsstrahlung emission is not expected at the considered time range. The exposure time of ICCD is $5 ns$ for the presented images in this report.  In other words, these experimental images correspond to plasma plume density in plane perpendicular to the magnetic field. The middle row shows the pseudo color plots of the plasma density in the cut plane perpendicular to the magnetic field and passing through the center of the plume from simulation. The bottom row the pseudo color plot is obtained after adding up the density of all the planes perpendicular to the magnetic field of the 3D plume image. This should approximately correspond to the images captured by the ICCD.  It can be seen that various aspects of the experimentally observed features are well captured in our simulations. The plume location as well as the dimensions of the diamagnetic cavity are accurately reproduced in simulations. Further, the simulation results show that once the cavity has attained maximum size it begins to collapse into a narrow jet like structure. The reduction in the size of the cavity is also found in experiment but jet like structure could not be observed in experiment. This difference could be attributed to the slight higher flow velocity of the plasma plume of the simulation as compared to the flow velocity of the plume observed in the experiment. Further, the shock front observed in the simulation images are sharper as compared to the experimental images because we have not included any dissipative processes like heat and electrical conduction in the simulation.

\begin{figure}[htbp]
	\centerline{\includegraphics[width=11cm, height=7cm]{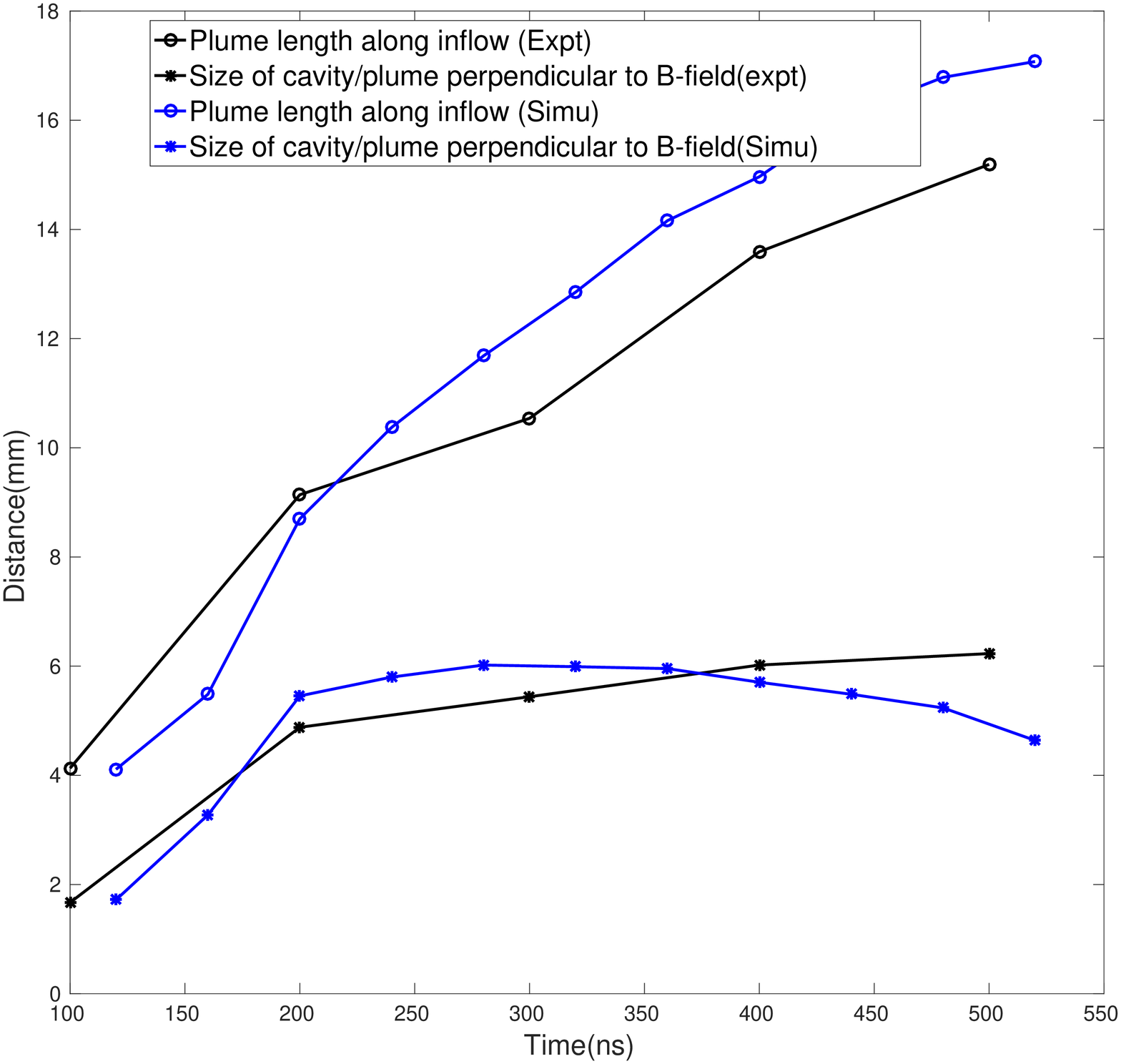}} %\centerline{\includegraphics[scale=0.5]{Figures/ComputationalDomain3D.eps}}
	\caption{Comparison of plume size as obtained from the experiment and that obtained from the simulation.}
	\label{Fig7_PlumeSizeExptSim}
\end{figure}

\begin{figure}[htbp]
	\centerline{\includegraphics[width=11cm, height=8cm]{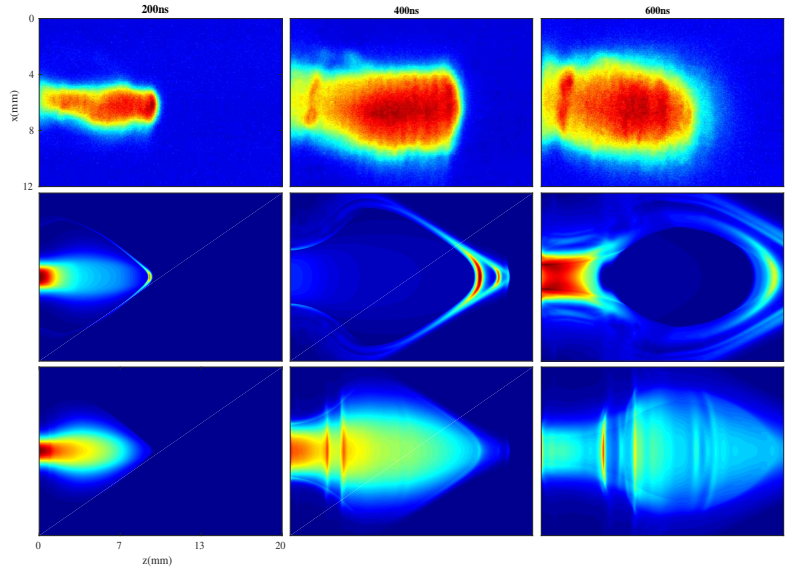}}
	\caption{Slice of plasma plume taken in a plane parallel to the direction of magnetic field(y-z plane).  The top row shows the images of plume at different times as obtained in the experiment.  The center row shows the plume density in the plane passing through the center  as obtained in simulation.The bottom row shows the plume density obtained after adding all the layers of the plume parallel to magnetic field as obtained in simulation.}
	\label{Fig8_Striations}
\end{figure}

In order to compare the plume size obtained in the experiment with the plume size obtained in the simulation, we have plotted the plume size along the direction of inflow $(\hat{z})$ and plume size in the direction $(\hat{x})$ perpendicular to both magnetic field and the inflow direction  as function of time for both simulation and experiment (Fig.\ref{Fig7_PlumeSizeExptSim}). The data from the experiment and simulation match quiet well. Fig.\ref{Fig8_Striations} shows the pseudo color plots of plasma plume density in the plane parallel to the external magnetic field obtained from simulation and experiment. The plasma plume expands along the magnetic field without much inhibition and the plasma density falls off drastically along this direction.This is evident if we compare the simulation images presented in Fig.\ref{Fig8_Striations} with the simulation images in Fig.\ref{Fig6_SimVsExpt}. However, such difference in size is not visible in the experimental images because the plasma density beyond the visible surface must have been so low that it is not detected by the ICCD. Further, at later times, the decelerating plasma plume is engulfed by the stretched magnetic field line and at same time there is mass density gradient at the surface of the plasma plume. This condition is ideal for  magnetic Rayleigh Taylor instability  to occur \cite{DiamagneticCavityInstabilityMHD1}. We believe that the striation like structures that appear in both simulation and experimental observations could be due the magnetic Rayleigh Taylor instability.  

\newpage
\section{Summary and conclusion}
We have presented a 3D ideal MHD simulation of laser generated plasma plume expansion in external transverse magnetic field and compared the results of the simulation with the experimental results. The simulation studies shows that the transverse magnetic field significantly governs the expansion of plasma plume and the observations made by this study have been found to compare reasonably well with the experimental observations.For instance, diamagnetic cavity formation and collapse of the plasma plume are the important experimental observations that are all observed in the ideal MHD simulation. Besides this, the plume size as function of time matches with that obtained in the experiment. The simulation results conclusively shows that all the major experimental observations could be attributed to the interplay between the $\overrightarrow{J} \times \overrightarrow{B}$ force and the force due thermal pressure. Furthermore, our simulation studies clearly demonstrate that the plasma plume expansion in the low magnetic field is reasonably well explained with the ideal MHD where the plasma based aspects like electrical resistivity, thermal diffusion, and Hall effect do not play any crucial role in the evolution of the plume.

\newpage

\section*{Acknowledgments}
 We thank  Dr.Mikhail Povarnitsyn for the VLL code and his guidance regarding using it. We thank Dr. Bhargav Viadya for useful discussions and suggestions regarding the PLUTO code. We acknowledge that the results reported in this article are the outcome of the simulations performed on the Antya cluster at Institute for Plasma Research.

\bibliographystyle{unsrt}
\bibliography{Bibliography}

\pagebreak%

\end{document}